\input{epsf}
\documentstyle[aps,amssymb,preprint]{revtex}
\begin{document}
\tightenlines

\title{Functional-integral based perturbation theory for the Malthus-Verhulst process}

\author{Nicholas R. Moloney$^{1,2,*}$ and Ronald Dickman$^{2,\dagger}$}

\address{$^1$ Institute of Theoretical Physics,
E\"otv\"os University, P\'azm\'any P\'eter s\'et\'any 1/A, \\
1117 Budapest, Hungary\\
$^2$Departamento de F\'{\i}sica, ICEx,
Universidade Federal de Minas Gerais,\\
30123-970
Belo Horizonte - Minas Gerais, Brasil\\
}

\date{\today}

\maketitle
\begin{abstract}
We apply a functional-integral formalism for Markovian birth and death
processes to determine asymptotic corrections to mean-field theory
in the Malthus-Verhulst process (MVP).  Expanding about the
stationary mean-field solution, we identify an expansion parameter
that is small in the limit of large mean population, and derive a
diagrammatic expansion in powers of this parameter.  The series is
evaluated to fifth order using computational enumeration of diagrams.
Although the MVP has no stationary state, we obtain good agreement
with the associated {\it quasi-stationary} values for the moments of
the population size, provided the mean population size is not small.
We compare our results with those of van Kampen's
$\Omega$-expansion, and apply our method to the MVP with input, for
which a stationary state does exist.
\end{abstract}
\vspace{1cm}

$^*$ email: moloney@general.elte.hu

$^\dagger$ email: dickman@fisica.ufmg.br

\newpage
\section{Introduction}

The need to analyze Markov processes described by a master equation
arises frequently in physics and related fields
\cite{vanK,gardiner}. Since such equations do not in general admit
an exact solution, approximation methods are of interest.  A widely
applied approximation scheme is van Kampen's `$\Omega$-expansion',
which furnishes corrections to the (deterministic) mean-field or
macroscopic description in the limit of large effective system size
\cite{vanK}. Another approximation method, based on a path-integral
representation for birth-and-death type processes, was proposed by
Peliti \cite{peliti85}. This approach was recently reviewed and
extended \cite{pert}, and applied to derive a series expansion for
the activity in a stochastic sandpile \cite{manpert}.

It should be noted that while effectively exact results can be
obtained via numerical analysis of the master equation, the
calculations become extremely cumbersome for large populations or
multivariate processes.  A further limitation of numerical analyses
is that they do not furnish algebraic expressions that may be
required in theoretical developments.  For these reasons it is
highly desirable to study approximation methods for stochastic
processes.

In the present work we apply the path-integral based perturbation approach to
a simpler problem, namely, the Malthus-Verhulst process (MVP), a birth-and-death
process in which unlimited population growth is prevented by a saturation effect.
(The death rate per individual grows linearly with population size.)  This is an
important, though highly simplified model in population dynamics.
Although the master equation for this process is readily solved numerically,
the model serves as a useful testing ground for approximation methods.
A lattice of coupled MVPs exhibits (in the infinite-size limit) a phase transition
belonging to the directed percolation universality class.

In the perturbation approach developed here \cite {peliti85,pert},
moments of the population size $n$ are expressed as functional
integrals over a pair of functions, $\psi(t)$ and $\tilde{\psi}(t)$,
involving an effective action.  The latter, obtained from an exact
mapping of the original Markov process, generally includes a part
that is bilinear in the functions $\psi(t)$ and $\tilde{\psi}(t)$,
whose moments can be determined exactly, and `interaction' terms of
higher order, that are treated in an approximate manner.  In the
present approach, the interaction terms are analyzed in a
perturbative fashion, leading to a diagrammatic series.  With
increasing order, the number of diagrams grows explosively, so that
it becomes convenient to devise a computational algorithm for their
enumeration and evaluation.  Elaboration of such an algorithm does
not, however, require any very sophisticated techniques, and could
in fact be applied to a variety of problems.  This is the approach
that was applied to the stochastic sandpile in Ref. \cite{manpert}.
In the latter case, the evaluation of diagrams involves calculating
multidimensional wave-vector integrals.  The present example is free
of this complication, allowing us to derive a slightly longer series
than for the sandpile.

In this work we focus on stationary moments of the MVP.  The series
expressions are apparently divergent, but nevertheless provide
nearly perfect predictions away from the small-population regime, as
compared with direct numerical evaluation of {\it quasi}-stationary
properties.   One might suppose that the divergent nature of the
perturbation series is due to the MVP not possessing a true
stationary state.  (The process must eventually become trapped in
the absorbing state, although the lifetime grows exponentially with
the mean population \cite{qss}.) Applying our method to the MVP with
a steady input, which does possess a stationary state, we find
however that the perturbation series continues to be divergent,
although again providing excellent predictions over most of
parameter space.

In the following section we define the Malthus-Verhlst process,
explain the perturbation method and report the series coefficients
for the first four moments, up to fifth order in the expansion
parameter.  In Sec. III we briefly compare these results to those of
the $\Omega$-expansion.  Then in Sec. IV we present numerical
comparisons of our method (and of the $\Omega$-expansions) against
quasi-stationary properties.  We apply our method to the MVP with
input in Sec. V, and summarize our findings in Sec. VI.

\section{Perturbation theory for the Malthus-Verhulst process}

Consider the Malthus-Verhulst process (MVP) $n(t)$, in
which each individual has a rate $\lambda$ to reproduce, and a rate
of $\mu + \nu (n-1)$ to die, if the total population is $n$.
By an appropriate choice of time scale we can eliminate one
of these parameters; we choose to set $\mu=1$.  Then in what follows we
use a dimensionless time variable $t' = \mu t$ and dimensionless
rates $\lambda' = \lambda/\mu$ and $\nu' = \nu/\mu$.  From here on we drop the
primes.

The mean-field or rate equation description of the process is
\begin{equation}
\dot{x} = (\lambda - 1)x - \nu x^2
\label{mve}
\end{equation}
where $x \equiv \langle n(t) \rangle$, with the nontrivial
stationary solution $\overline{x} = (\lambda - 1)/\nu$
for $\lambda > 1$.  We are interested in deriving systematic
corrections to this result, and in calculating higher moments
of the process.

Our starting point is the expression for the $r$-th factorial
moment of a general process taking non-negative integer values,
\begin{equation}
\langle n^r (t) \rangle_f = e^{-p}
U_t^{(r)} (\zeta \!=\! p) \;,
\label{fmpoi}
\end{equation}
where, for simplicity, we have assumed an initial Poisson
distribution with parameter $p$ (see Eq. (108) of Ref. \cite{pert}),
so that the probability generating function at time zero is
$\Phi_0(z) = e^{p(z-1)}$. Here, the kernel $U_t^{(r)}$ is given by
the functional integral,
\begin{equation}
U_t^{(r)} (\zeta) \equiv \left(\! \frac{\partial^r U_t
(z,\zeta)}{\partial z^r} \! \right)_{z=1} = \int  \!{\cal D} \!
{\psi} \! \int \! {\cal D} \hat{\psi} \, \psi(t)^r {\cal F}
[\psi,\hat{\psi}]_{z=1} \exp[-S_I] \;, \label{drUdzr}
\end{equation}
(equivalent to Eq. (106) of \cite{pert}), with
\begin{equation}
{\cal F}[\psi,\hat{\psi}]_{z=1} =
 \exp \!\left[\! -\! \int_0^t\! \!dt'
\hat{\psi}[\partial_{t'} \!+\! (1 \!-\! \lambda)]\psi
+\! \zeta \right]
\label{F}
\end{equation}
containing the bilinear part of the action, and, in the case of the
MVP, the ``interaction" part,
\begin{equation}
S_I = \int_0^t dt'[- \lambda \hat{\psi}^2 \psi
\!+\! \nu \hat{\psi}(1\!+\!\hat{\psi}) \psi^2 ]
\equiv \int_0^t dt'{\cal L}_I (t'),
\label{si}
\end{equation}
(see Eq. (54) of \cite{pert}).

Our goal is to expand each factorial moment about its mean-field
value.  To this end, consider the shift of variable,
\begin{equation}
\psi(t) = \overline{n} + \phi(t)
\label{shift}
\end{equation}
where $\overline{n}$ is a real constant.
On performing this shift, the argument of the exponential
(i.e., of the factors ${\cal F}[\psi,\hat{\psi}]_{z=1}$ and
$\exp[-S_I]$) in Eq. (\ref{drUdzr}) becomes,
\begin{eqnarray}
\nonumber
\zeta &+& \int_0^t dt'\{
-\hat{\psi}[\partial_{t'} \!+\! 1 \!-\! \lambda + 2\nu \overline{n}]\phi
-\overline{n}(1 \!-\! \lambda \!+\! \nu \overline{n})\hat{\psi}
\\
&+& \overline{n}(\lambda - \nu \overline{n}) \hat{\psi}^2
+ (\lambda - 2\nu \overline{n}) \hat{\psi}^2 \phi
- \nu  \hat{\psi} \phi^2 - \nu \hat{\psi}^2 \phi^2 \}
\label{shift1}
\end{eqnarray}
This simplifies if we
let $\overline{n} = (\lambda - 1)/\nu$, which eliminates the
term $\propto  \hat{\psi}$.  Introducing
$w \equiv \lambda \!-\! 1$ (equal to $-w$ as defined
in \cite{pert}), the argument of the exponential is:
\begin{eqnarray}
\nonumber
\zeta &+& \int_0^t dt'\{
-\hat{\psi}[\partial_{t'} \!+\! w]\phi
+ \overline{n} \hat{\psi}^2
\\
&+& (2 \! -\! \lambda) \hat{\psi}^2 \phi
- \nu  \hat{\psi} \phi^2 - \nu \hat{\psi}^2 \phi^2 \}
\label{shift2}
\end{eqnarray}
We recognize the exponential of $\zeta$ plus the first term in the integrand as
${\cal F}[\hat{\psi},\phi]_{z=1}$; the remaining terms then
represent $-S_I'$, the new effective interaction, which
will be treated perturbatively.  The four terms of $S_I'$ are
represented graphically, following the conventions of Ref. \cite{pert},
in Fig. 1.

\begin{figure}[b]
\epsfysize=3cm
\epsfxsize=6cm
\centerline{
\epsfbox{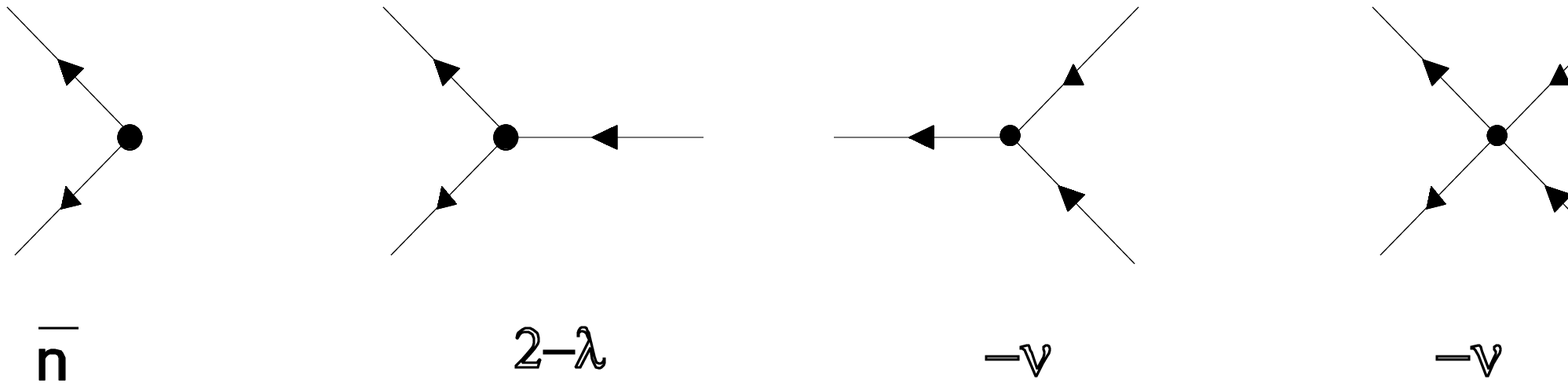}}
\label{c1f1}
\end{figure}
\begin{center}
{\sf Fig.1. Vertices in the perturbation series for the MVP.}
\end{center}

\noindent We refer to these vertices as source, bifurcation, conjunction and
4-vertex, respectively.

\subsection{Perturbation expansion}

The analysis of the MVP now follows the lines of \cite{pert}. Let
\begin{equation}
[{\cal A}] \equiv  \int  \!{\cal D} \! {\phi} \! \int \! {\cal D}
\hat{\psi} \, {\cal A }(\phi,\hat{\psi}) {\cal F}[\phi,\hat{\psi}],
\label{expect}
\end{equation}
denote the {\it free expectation} of any function ${\cal A}$ of
$\phi$ and $\hat{\psi}$. From the discussion of Sec. 4 Ref.
\cite{pert} we have,
\begin{equation}
[\phi(t)^r] = (p-\overline{n})^r e^{-rwt},
\label{expphi}
\end{equation}
(here we used $\phi(0) = p - \overline{n}$, and have already
canceled the factor $e^\zeta$ in $U_t^0$ with the corresponding
factor in the inital generating function, $\Phi_0(\zeta)$),
\begin{equation}
[\hat{\psi}(t)] = 0,
\label{exppsihat}
\end{equation}
and
\begin{equation}
[\phi(t_1) \hat{\psi}(t_2)] = \Theta (t_1 - t_2) e^{-w(t_1 - t_2)}.
\label{prop}
\end{equation}

Consider now the series for the mean population size $\langle n(t) \rangle$.
The zeroth-order term is simply
\begin{equation}
\langle n(t) \rangle_0 = \overline{n} + [\phi(t)]
= \overline{n} + (p-\overline{n}) e^{-wt}
\label{nzero}
\end{equation}
which converges to the mean-field result as $t \to \infty$.
Similarly, the leading term in $\langle n^r (t) \rangle_f$ is
$\overline{n}^r$, so that the stationary probability distribution,
in mean-field approximation, is Poisson with parameter
$\overline{n}$.

Corrections involving $S_I'$ are conveniently represented as diagrams
with all lines leaving vertices contracted with ingoing lines,
either at vertices to the left, or at the ``sink" lying to the left of
all vertices.  (In the calculation of $\langle n^r (t) \rangle_f$ the sink
is a point with $r$ incoming lines.)
The lowest-order diagram in $\langle n (t) \rangle$ is depicted in Fig. 2.

\begin{figure}[h]
\epsfysize=3cm
\epsfxsize=6cm
\centerline{
\epsfbox{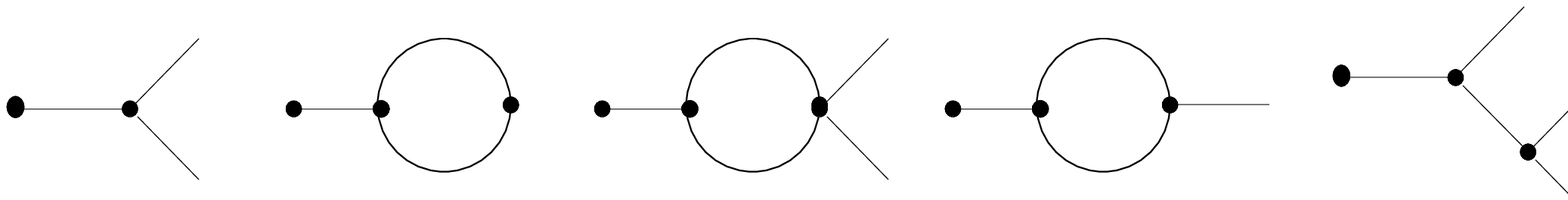}}
\label{c1f2}
\end{figure}
\begin{center}
{\sf Fig. 2. One- and two-vertex diagrams in the series for $\langle n(t) \rangle$.}
\end{center}

\noindent This diagram makes the following contribution to $\langle n(t) \rangle$:
\begin{equation}
- \nu \int_0^t dt_1 e^{-w(t-t_1)} (p-\overline{n})^2 e^{-2w t_1}
= - \frac{\nu}{w} (p-\overline{n})^2 e^{-wt} (1 - e^{-wt}).
\label{n1}
\end{equation}
The four diagrams (in the series for $\langle n(t) \rangle$)
having two vertices are also shown in Fig. 2.  Only the
first of these four makes a nonzero contribution to
$\langle n_\infty \rangle \equiv \lim_{t \to \infty} \langle n(t) \rangle$,
namely,
\begin{equation}
- 2\nu  \overline{n} \int_0^t dt_1 \int_0^{t_1} dt_2 e^{-w(t-t_1)}  e^{-2w (t_1-t_2)}
= - \frac{1}{w} (1 - e^{-wt})^2.
\label{ninf2}
\end{equation}
(The combinatorial factor 2 represents the number of ways the lines may be
contracted between source and bifurcation.)

If we are only interested in stationary properties, it is convenient to use the
Laplace transform.  Let $f_D$ be the $n$-fold integral over time variables
in a given $n$-vertex diagram $D$.  Noting that time-dependent factors are associated with
each propagator and each uncontracted incoming line, we see that $f_D$ is of the
form:
\begin{equation}
\label{fD}
f_D(t) = \int_{0}^{t} \! dt_{1}\int_{0}^{t_{1}} \! dt_{2}\ldots
\int_{0}^{t_{n-1}} \!dt_{n} \,e^{-\alpha_{1}(t-t_{1})-\alpha_{2}(t_{1}-t_{2})\ldots-
\alpha_{n}(t_{n-1}-t_{n})-\beta_1 t_1 -\cdots -\beta_n t_n}
\end{equation}
where $\beta_i$ is $w$ times the number of uncontracted lines incident on vertex $i$.
(In the first graph of Fig. 2, $\beta_1 = 2w$.)  The factors $\alpha_i$ are given
by $w$ times the number of lines (propagators) running between vertices $i$ and
$i\!-\!1$ (with $i=0$ representing the sink), regardless of where these lines
originate or terminate.  Now consider
\begin{equation}
\label{fDs}
\tilde{f}_D (s) = \int_{0}^\infty dt e^{-st} f_D(t) .
\end{equation}
Using Eq. (\ref{fD}) we may write
\begin{eqnarray}
\nonumber
\tilde{f}_D (s) &=&
\int_{t_{1}}^{\infty} \! dt \int_{t_{2}}^{\infty} \! dt_{1}\ldots\int_{0}^{\infty} \! dt_{n}
\exp[-(\alpha_{1}+s) (t \!-\! t_{1}) - (\alpha_{2}+\beta_1+s)(t_1 \! - \! t_2)
\\
\nonumber
&\;& \;\;\;\;- (\alpha_3 \!+\! \beta_1\!+\! \beta_2 \!+\! s)(t_2 \!-\! t_3) \!-\! \cdots
\!-\! (\alpha_n \!+\! \beta_1 \!+\! \cdots \!+\! \beta_{n-1} \!+\! s)(t_{n-1} \!-\! t_{n})
\\
\nonumber
&\;& \;\;\;\;- (\beta_1 + \cdots + \beta_n +s) t_{n} ]
\\
& = & \left[(\alpha_{1} \!+\! s)(\alpha_{2} \!+\! \beta_1 \!+\! s) \cdots
(\alpha_n \!+\! \beta_1 \!+\! \cdots \!+\! \beta_{n-1} \!+\!s)
(\beta_1 \!+\! \cdots \!+\! \beta_n \!+\! s)
\right]^{-1} .
\label{fDs2}
\end{eqnarray}
The contribution of $D$ to $\langle n_\infty \rangle $ is proportional to
\begin{equation}
\overline{f}_D \equiv \lim_{t \rightarrow \infty} f_D(t)
= \lim_{s \to 0} s \tilde{f}_D(s),
\label{fbar}
\end{equation}
by the Final Value Theorem of Laplace transforms.
This is {\it zero} unless $\beta_i = 0$,
$\forall i = 1,...,n$.  In the latter case,
\begin{equation}
\overline{f}_D = \left[(\alpha_{1}+s)(\alpha_{2}+s) \cdots (\alpha_n + s) \right]^{-1} .
\label{fDs3}
\end{equation}
Thus the only diagrams contributing to $\langle n_\infty \rangle $ (and by extension,
to $\langle n_\infty^r \rangle $) are those free of uncontracted lines.  This is in fact
evident on physical grounds: each such line carries a factor $p-\overline{n}$, whereas
stationary properties cannot depend on the initial mean population $p$.
In diagrams contributing to  $\langle n_\infty \rangle $, the first vertex
(i.e., immediately to the right of the sink) must be a conjunction, while the
$n$-the vertex must be a source.

\subsection{Diagrammatic analysis for the stationary mean population}

The contribution of a diagram $D$ to $\langle n_\infty \rangle $ is the product
of three factors: $\overline{f}_D$ discussed above; a combinatorial factor
counting the number of contractions consistent with the diagram topology;
the product of vertex-associated factors shown in Fig. 1.

Certain infinite classes of diagrams may be summed up exactly.
Consider the sequence shown in Fig. 3: between the source and
conjunction we insert any number of 4-vertices or
bifurcation-conjunction pairs.

\begin{figure}[h]
\epsfysize=2cm
\epsfxsize=6cm
\centerline{
\epsfbox{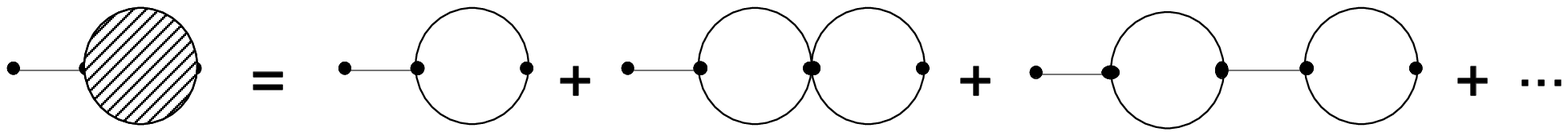}}
\label{c1f3}
\end{figure}
\begin{center}
{\sf Fig. 3. A summable set of diagrams in the series for $\langle n(t) \rangle$.}
\end{center}

\noindent A short calculation shows that the contribution of this series
is given by:
\begin{equation}
-\frac{\overline{n} \nu}{w^2} \sum_{n=0}^\infty \left[ \left( -\frac{\nu}{w} \right)
\left( 1+ \frac{2 \!-\! \lambda}{w} \right) \right]^n
= - \frac{1}{w+\nu /w} .
\label{drloop}
\end{equation}
In this way, by multiplying the $n=2$ contribution, $ - \frac{1}{w}$,
by $\kappa \equiv (1 + \nu /w^2)^{-1}$, we have included all diagrams of the form
of Fig. 3, i.e., diagrams in which all lines exiting vertex $i$ are
contracted on vertex $i\!-\!1$, $\forall i = 1,...,n$.  With this simple replacement
such {\it reducible} diagrams
are no longer to be included explicitly.

A similar observation allows us to add arbitrary sequences of
4-vertices or bifurcation-conjunction pairs to any diagram
of four or more vertices.  (Diagrams outside the series depicted in
Fig. 3 have $n \geq 4 $ vertices.)  The diagram consists of a `body'
containing vertices $2,...,n\!-\!1$ linked to a conjunction (vertex
1) and a source (vertex $n$).  Call the factor associated with the
body $A$, so that the diagram makes a contribution of $-\nu A
\overline{n} $ to $\langle n_\infty \rangle $. A family of diagrams
may be constructed by adding sequences, as above, between
vertex $n$ and the body; the sum of these contributions is
\begin{equation}
-\nu A \overline{n} \sum_{n=0}^\infty \left( -\frac{\nu}{w^2} \right)^n
= - \nu A \overline{n} \frac{1}{1+\nu/w^2} .
\label{drsource}
\end{equation}
By the same reasoning we may insert arbitrary sequences between the first
vertex and the body, yielding the same set of contributions.
As a result, we may multiply the contribution of any diagram (not included
in the sequence of Fig. 3) by $\kappa^2$, and thereby
automatically include all diagrams with the same body but arbitrary
linear sequences between the first vertex and the body, and between
vertex $n$ and the body.  Such diagrams are no longer included explicitly.
This procedure will be called ``dressing" the conjunction (vertex 1)
and the source (vertex $n$).

At this stage we have essentially exhausted the simplifications (via
``dressing" or summing sets of trivial variations to a diagram) that can
be realized in the Laplace transform representation.  Summarizing, the
contributions to $\langle n_\infty \rangle $ are as follows.

\noindent 1. The mean-field contribution $\overline{n}$.

\noindent 2. The set of diagrams shown in Fig. 3, giving
$ - \kappa / w$.

\noindent 3. Diagrams of four or more vertices, with no dangling lines,
and subject to the restrictions mentioned above.  Each such contribution
is to be multiplied by $\kappa^2$.

\noindent Including diagrams of up to four vertices we find:

\begin{eqnarray}
\nonumber
\langle n_\infty \rangle &=& \overline{n} \left[ 1 - \frac{\kappa \nu}{w^2}
+ \frac{2\kappa^2  \nu^2 (2 - \lambda)}{w^4} + \cdots \right]
\\
&=& \overline{n} \left[ 1 - (1- \kappa)
+ 2(2 - \lambda) (1- \kappa)^2 + \cdots \right]
\label{exp4}
\end{eqnarray}
where the higher order terms come from diagrams of five or more vertices.
This result suggests that we adopt
\begin{equation}
\epsilon \equiv 1- \kappa = \frac{\nu}{w^2 + \nu}
\label{defepsilon}
\end{equation}
as the expansion parameter.  Up to diagrams of four vertices we have
\begin{equation}
\langle n_\infty \rangle =
\overline{n} \left[ 1 - \epsilon
+ 2(2 - \lambda) \epsilon^2 + \cdots \right].
\label{exp4v}
\end{equation}
The first correction $-\overline{n} \epsilon \propto 1/\lambda$,
as $\lambda \to \infty$.

Note that $\nu \propto \epsilon$ to lowest order.
The lowest order in $\epsilon$ at which a given diagram contributes
to $\langle n_\infty \rangle$, when expressed in the form
of Eq. (\ref{exp4v}), may be found as follows.  First observe that
the diagram carries a factor $\nu^{c+f-s+1}$ where $c$, $f$ and $s$
represent the number of conjunctions, four-vertices and sources, respectively,
and the contribution of 1 in the exponent is due to the prefactor
$\overline{n}$.  Certain constraints exist between the numbers of
vertices.  Equating the total number of lines emanating from
vertices with the total number entering, we find
\begin{equation}
2(s + f + b) + c = 2(f+c) + b + 1
\label{vrel1}
\end{equation}
where $b$ is the number of bifurcations and the 1 on the r.h.s. represents
the sink.  Thus $c = 2s + b - 1$.  On the other hand the total number of
vertices is $n = s + b + f + c$, and using this we
find the power of $-\nu$ to be $c+f-s+1 = n-c \equiv m$.

A simple analysis yields the algebraic factor associated with a given
diagram having $n \geq 4$ vertices:
\begin{equation}
F_{alg} = (-1)^{c+f} \kappa^2 (2- \lambda)^b w^{s-n-1} \nu^{n-c}.
\label{falg}
\end{equation}
This is readily expressed in terms of the parameter $\epsilon$ as:
\begin{equation}
F_{alg} = (-1)^{c+f} (2 - \lambda)^b w^f
 \frac{\epsilon^m}{(1-\epsilon)^{m-2}}.
\label{falgep}
\end{equation}

Since we intend to organize the expansion in powers of $\epsilon$, it is
of interest to know which values of $n$ correspond to a given $m$.
We begin by noting that in diagrams of four or more vertices, the
second vertex (from the left) must be a conjunction, just as the first is.
(If it were a bifurcation or a four-vertex the result would be a diagram
already included by dressing the first vertex.)  Thus $c \geq 2$, and there
are in fact diagrams with just two conjunctions, for {\it any} $n \geq 4$.
We therefore have $m \leq n-2$ or $n \geq m+2$.  To find an upper bound on
$n$, we find an upper bound on $c$.  Recall that $c = 2s + b - 1$.  To
maximize $c$ we let $b=f=0$ (a diagram consisting of only sources
and conjunctions), in which case $c+s = n$, yielding
$m = n-c = s = (n+1)/3$, or $n \leq 3m-1$.  Thus for
$m=2$ we have contributions from diagrams of 4 and 5 vertices,
while for $m=5$ the number of vertices ranges from 7 to 14.

In order to evaluate the contributions from diagrams of five or more
vertices we have developed an enumeration code.  Reducible diagrams
are eliminated by imposing the following rules:

1) The second vertex, like the first, is a conjunction.

2) Vertex $n-1$ cannot be a 4-vertex.

3) If vertex $n-1$ is a conjunction, vertex $n-2$ must be a source.

Using the enumeration code we are able to evaluate contributions up
to ${\cal O}(\epsilon^5)$.  At this order there are approximately
$8 \times 10^8$ diagrams.  (There are 9, 1317 and 594$\,$339 diagrams at orders
2, 3 and 4, respectively.)  This explosive growth in the number of terms
prevents our going to higher order.
The fifth-order calculation requires about two days on a
fast PC.

Writing Eq. (\ref{exp4v}) in the form
$\langle n_\infty \rangle = \overline{n} [1 + \sum_{m \geq 1} g_m \epsilon^m]$,
and using $u\equiv 2-\lambda$, we have:
\begin{eqnarray}
\nonumber
g_1 &=& -1,
\\
\nonumber
g_2 &=& 2u - 5,
\\
\nonumber
g_3 &=& -10 u^2 + (55-4w)u + 19w - 60,
\\
\nonumber
\\
\nonumber
g_4 &=&
78 u^3 -(689-62w)u^2 +(1665-504w+8w^2)u
\\
\nonumber
&\,& \; -65w^2 +745w -1165,
\\
\nonumber
\\
\nonumber
g_5 &=& -750u^4 + (9437-880w)u^3 -(37078-10259w+266w^2)u^2
\\
\nonumber
&\,& \; +(57620-32328w+3117w^2-16w^3)u
\\
&\,& \; + 211w^3 -6040w^2 +27786w -29390
\label{gm}
\end{eqnarray}

A particularly simple case is $\lambda=2$, for which we have

\begin{equation}
\langle n \rangle =
\frac{1}{\nu}[1 - \epsilon - 5\epsilon^2 - 41\epsilon^3 -
485\epsilon^4 - 7443\epsilon^5 -...]
\label{lambda2}
\end{equation}
The ratios of successive coefficients in the series are:
1, 5, 8.2, 11.829 and 15.346, suggesting unlimited growth
and therefore a divergent series.  Numerical results (see below)
nevertheless reveal excellent agreement with quasi-stationary properties
for $\lambda$ sufficiently large.

\subsection{Higher moments}

We turn now to the expansion of higher stationary factorial moments.
From Eqs. (\ref{drUdzr}) and (\ref{shift}) we have
\begin{equation}
\langle n^r \rangle_f = [(\overline{n} + \phi)^r e^{-S_I}] .
\label{rthmom}
\end{equation}
Expanding the product, we have first the mean-field contribution $\overline{n}^r$,
and then a series of terms involving diagrams.  The term $\propto [\phi^q e^{-S_I}]$
involves (for $q < r$) diagrams that already appeared in the series for the $q$-th
factorial moment.  Thus for $r=2$ we have
\begin{equation}
\langle n^2 \rangle_f = \overline{n}^2 + 2 \overline{n}[ \phi e^{-S_I}]
+[ \phi^2 e^{-S_I}].
\label{mom2}
\end{equation}
Note that $[ \phi e^{-S_I}]$ is simply ${\cal D}_1$, the sum of all diagrams with a
single line entering the sink, that is, the diagrammatic series for
$\langle n \rangle$.  Consider now the final term in Eq. (\ref{mom2}), the series
${\cal D}_2$ of diagrams with two lines entering the sink.  Recalling that, in all
diagrams in ${\cal D}_1$, the first (leftmost) vertex is a conjunction, we see that
there is a one-to-one correspondence between ${\cal D}_1$ and ${\cal D}_2$.
For a given diagram, making a contribution of $A$ to ${\cal D}_1$,
there is a corresponding diagram (with the conjunction and one-line sink
replaced by a two-line sink) that contributes $-wA/\nu = - \overline{n} A$ to
${\cal D}_2$, that is, ${\cal D}_2 = - \overline{n}{\cal D}_1$.  Thus we have
\begin{eqnarray}
\nonumber
\langle n^2 \rangle_f &=& \overline{n}^2 + \overline{n} {\cal D}_1
\\
&=& \overline{n} \langle n \rangle,
\label{mom2a}
\end{eqnarray}
where in the second line we used $\langle n \rangle = \overline{n} + {\cal D}_1$.
Using Eqs. (\ref{gm}) and (\ref{mom2a}), we find that
\begin{equation}
\mbox{var}(n) - \langle n \rangle =
\overline{n}^2 \epsilon [1 - (2u-4) \epsilon + \cdots].
\label{poidev}
\end{equation}
For a Poisson distibrubtion the difference is zero; here it approaches
$1/\nu$ as $\lambda \to \infty$.

For $r \geq 3$ we do not have a simple relation between $D_r$ and $D_1$, so the
diagrammatic series must be evaluated to the desired order.  In general we have
\begin{equation}
\langle n^r \rangle_f = \sum_{j=0}^r
\left(
\begin{array}{c}
r
\\
j
\end{array}
\right)
\overline{n}^{r-j} {\cal D}_j ,
\label{rthmoma}
\end{equation}
where ${\cal D}_1 = [\phi^j e^{-S_I}]$, with ${\cal D}_0 = 1$.
(Note that for $r=3$ the $j=1$ and $j=2$ terms cancel.)  As in the
case of $\langle n \rangle$, we include an overall factor of $\overline{n}^r$,
and write
\begin{equation}
\langle n^r \rangle_f = \overline{n}^r \left[1 +
\sum_{j=1}^r
\left(
\begin{array}{c}
r
\\
j
\end{array}
\right)
\left(\frac{\nu}{w} \right)^j {\cal D}_j \right].
\label{rthmomb}
\end{equation}
Note that for $r \geq 3$ we can dress the leftmost source, as before,
but that it is no longer possible to dress the sink.  Thus
the algebraic factor associated with an arbitrary diagram in
$(\nu/w)^r {\cal D}_r$ is
\begin{equation}
F_{alg} = (-1)^{c+f} \kappa u^b w^{s-n-r} \nu^{c+f-s+r}.
\label{falga}
\end{equation}
An analysis along the lines presented in the $r=1$ case leads to
$m = n-c$ (as before) for the order in $\nu$, and to $s-n-r = f$,
so that
\begin{equation}
F_{alg} = (-1)^{c+f} u^b w^f \frac{\epsilon^m}{(1-\epsilon)^{m-1}}.
\label{falgb}
\end{equation}
The limits on $n$, for fixed order $m$, are $m \leq n \leq 3m - r$.
(Note that $n=m$ is possible for $r \geq 3$ because there are
diagrams with no conjunctions.)

As in the case $r=1$, we have constructed an enumeration code to evaluate the
corrections due to diagrams.  Writing the contribution to
$\langle n^r \rangle_f$ coming from
${\cal D}_r$ as $\overline{n}^r \sum_{m \geq 1} g_{r,m} \epsilon^m $,
we have, to $r=4$ and $m=5$:
\begin{eqnarray}
\nonumber
g_{3,1} &=& 0,
\\
\nonumber
g_{3,2} &=& 2u - 5,
\\
\nonumber
g_{3,3} &=& -10 u^2 + (57-4w)u + 19w - 65,
\\
\nonumber
\\
\nonumber
g_{3,4} &=&
78 u^3 -(699-62w)u^2 +(1722-508w+8w^2)u
\\
\nonumber
&\,& \; -65w^2 +764w -1230,
\\
\nonumber
\\
\nonumber
g_{3,5} &=& -750u^4 + (9515-880w)u^3 -(37777-10581w+266w^2)u^2
\\
\nonumber
&\,& \; +(58432-32836w+3125w^2-16w^3)u
\\
&\,& \; -30560 + 28550 w -6105w^2 +211 w^3
\label{gmr3}
\end{eqnarray}
and
\begin{eqnarray}
\nonumber
g_{4,1} &=& 0,
\\
\nonumber
g_{4,2} &=& 3,
\\
\nonumber
g_{4,3} &=& 6 u^2 -41 u -9 w +53,
\\
\nonumber
\\
\nonumber
g_{4,4} &=&
-54 u^3 + (547-30w)u^2 -(1452-330 w)u
\\
\nonumber
&\,& \; +27 w^2 - 576w +1088,
\\
\nonumber
\\
\nonumber
g_{4,5} &=& 570 u^4 - (7785-564w)u^3 +(32515-7799w+114w^2)u^2
\\
\nonumber
&\,& \; -(52118-26354w+1895w^2)u
\\
&\,& \; + 28058 - 24234 w +4361w^2 -81 w^3.
\label{gmr4}
\end{eqnarray}
Numerical comparisons of these series against quasi-stationary
properties of the MVP will be discussed in Sec. 4.

\section{Comparison with the $\Omega$-expansion}

A well known method for obtaining approximate solutions to the
master equation is van Kampen's $\Omega$-expansion \cite{vanK}. The
method depends on the system size (or the expected value of the
stochastic variable of interest) being large, so that fluctuations
are small compared to the value $\overline{n}(t)$ predicted by the
macroscopic equation. The stochastic variable $n$ is then written in
the form

\begin{equation}
n = \Omega \zeta(t) + \Omega^{1/2} \xi(t)
\label{Omega1}
\end{equation}
where the first, deterministic, term represents the solution to the
macroscopic equation, with $\Omega$ denoting the size of the system.
The stochastic contribution, represented by the second term, is
expected on general grounds to scale as $\Omega^{1/2}$.  As shown in
Ref. \cite{vanK}, $\zeta(t)$ satisfies the macroscopic equation,
while the probability density $\Pi (\xi,t)$ satisfies, to lowest
order in $\Omega^{-1/2}$, a linear Fokker-Planck equation.  In the
present case the macroscopic equation is

\begin{equation}
\frac{d\zeta}{dt} = (\lambda - 1) \zeta - \nu \zeta^2,
\label{macro}
\end{equation}
i.e., the Malthus-Verhulst equation, with nontrivial stationary
solution $\zeta = (\lambda - 1)/\nu = \overline{n}$.
The equation governing the evolution of $\Pi$ is, in the
present instance,

\begin{equation}
\frac{\partial \Pi}{\partial t} = (1 + 2\nu \zeta - \lambda)
\frac{\partial}{\partial \xi} (\xi \Pi) + \frac{1}{2} \zeta (1 + \nu
\zeta + \lambda) \frac{\partial^2 \Pi}{\partial \xi^2}. \label{fpe}
\end{equation}
This implies that in the stationary state, $\xi$ is Gaussian with
mean zero and variance $\lambda/\nu$, so that, setting the formal
expansion parameter $\Omega$ to unity in Eq. (\ref{Omega1}), $n$ is
Gaussian with mean $\overline{n}$ and variance $\lambda/\nu$. In the
perturbation expansion developed in the preceding section, $n$ is,
to zeroth order, a Poissonian random variable with mean
$\overline{n}$.  Conceptually, a Poisson distribution seems
preferable to a Gaussian as the reference distribution (since $n$ is
discrete and cannot assume negative values), but in practical terms
we expect the difference to be very small.

To first order in $\epsilon$, we found $\langle n \rangle =
\overline{n}(1-\epsilon)$ and var($n) =
\overline{n}(1+\overline{n}\epsilon)$.  Noting that
\begin{equation}
\overline{n}(1+\overline{n}\epsilon) = \frac{\lambda-1}{\nu} +
\left(\frac{\lambda-1}{\nu}\right)^2 \frac{\nu}{(\lambda-1)^2 + \nu}
= \frac{\lambda}{\nu} - \frac{1}{(\lambda-1)^2},
\label{diffvar}
\end{equation}
we see that the difference from the $\Omega$-expansion result, to
first order in $\Omega^{-1/2}$, is small for $\lambda \gg 1$.  In
the $\Omega$-expansion, corrections to the moments $\langle \xi^r
\rangle$ can be obtained via the procedure detailed in Ref.
\cite{vanK}.  In the present case one finds
\begin{equation}
\langle \xi \rangle = -\frac{1}{\lambda-1} \Omega^{-1/2},
\label{xiomega}
\end{equation}
so that
\begin{equation}
\langle n \rangle = \overline{n} \left[ 1 -
\frac{\nu}{(\lambda-1)^2} \right]
\label{nomega}.
\end{equation}
The series prediction is
\begin{equation}
\langle n \rangle = \overline{n} ( 1 - \epsilon) = \overline{n}
\left[ 1 - \frac{\nu}{(\lambda-1)^2} + {\cal O} (\epsilon^2)
\right],
\label{nepsilon}
\end{equation}
so that the two methods agree to first order in $\epsilon$.
Extending the $\Omega$-expansion to include terms of order
$\Omega^{-1}$ in $\xi$, one finds
\begin{equation}
\langle n \rangle = \overline{n} \left[ 1 -
\frac{\nu}{(\lambda-1)^2} - \frac{\nu^2}{(\lambda-1)^4}(\lambda^2 -
3\lambda + 4) \right]. 
\label{nomega2}
\end{equation}
These results are compared against the $\epsilon$ series in the
following section.

\section{Numerical Comparisons}

In this section we compare the $\epsilon$-series predictions with
exact (numerical) results for quasi-stationary (QS) properties of
the MVP.  The latter are obtained via recursion relations leading to
the QS distribution as detailed in \cite{qss}.  QS properties are
those obtaining at arbitrarily long times, conditioned on survival
(i.e., the process has never visited the absorbing state). Although
a condition on survival is not involved in the perturbative analysis
of Sec. 2, it seems reasonable to compare its predictions with QS
properties, since the true stationary properties are the trivial
ones of the absorbing state (population zero, no fluctuations). We
note that, as suggested above, the series in $\epsilon$ appears to
be divergent for any set of parameter values, as reflected in the
ratios between coefficients of successive terms, whose ratios appear
to grow without limit. For suitably large values of $\lambda$ the
ratios are very small, although increasing with order $m$, so that
the series, truncated at fifth order, appears to have ``converged".
The numerical evidence (limited to $m \leq 5$, of course) points
nevertheless to a divergent series.

In Fig. 4 we compare the QS mean population size with $\langle n_\infty \rangle$
as predicted by the series, as a function of $\lambda$, with $\nu$ fixed at
0.01.  For $\lambda $ less than about 1.2, the series is useless (it yields a
negative population size!), and is inferior to mean-field theory.  For $\lambda \geq 1.4$
on the other hand, we find good agreement with the QS result.  For a more
detailed comparison we plot, in Fig. 5, the difference $\Delta n$ between
$\langle n \rangle$ (as given by the QS distribution and the $\epsilon$ series)
and the mean-field prediction $\overline{n}$; there is perfect agreement for
$\lambda \geq 1.5$.  (The correction to the mean population size decays
$\propto 1/\lambda$ for large $\lambda$.  This is readily seen from
Eq. (\ref{exp4}) if we note that $\overline{n} \epsilon \sim 1/\lambda$
for $\lambda \gg 1$.)  Figures 6 and 7 show that similar
behavior obtains for $\nu = 0.1$ and 0.001.  The smaller $\nu$, the
smaller the value of $\lambda$ required for
agreement between series and QS values.   (For $\nu = 0.1$
agreement sets in for $\epsilon \leq 0.024$, approximately;
in the other cases $\epsilon \leq 0.06$ is sufficient.)
At the point where the series and QS results begin to agree, the
mean population is not particularly large; for $\nu = 0.01$,
$\lambda = 1.4$ corresponds to $\langle n \rangle \simeq 35$.

\begin{figure}[h]
\epsfysize=6cm
\epsfxsize=6cm
\centerline{
\epsfbox{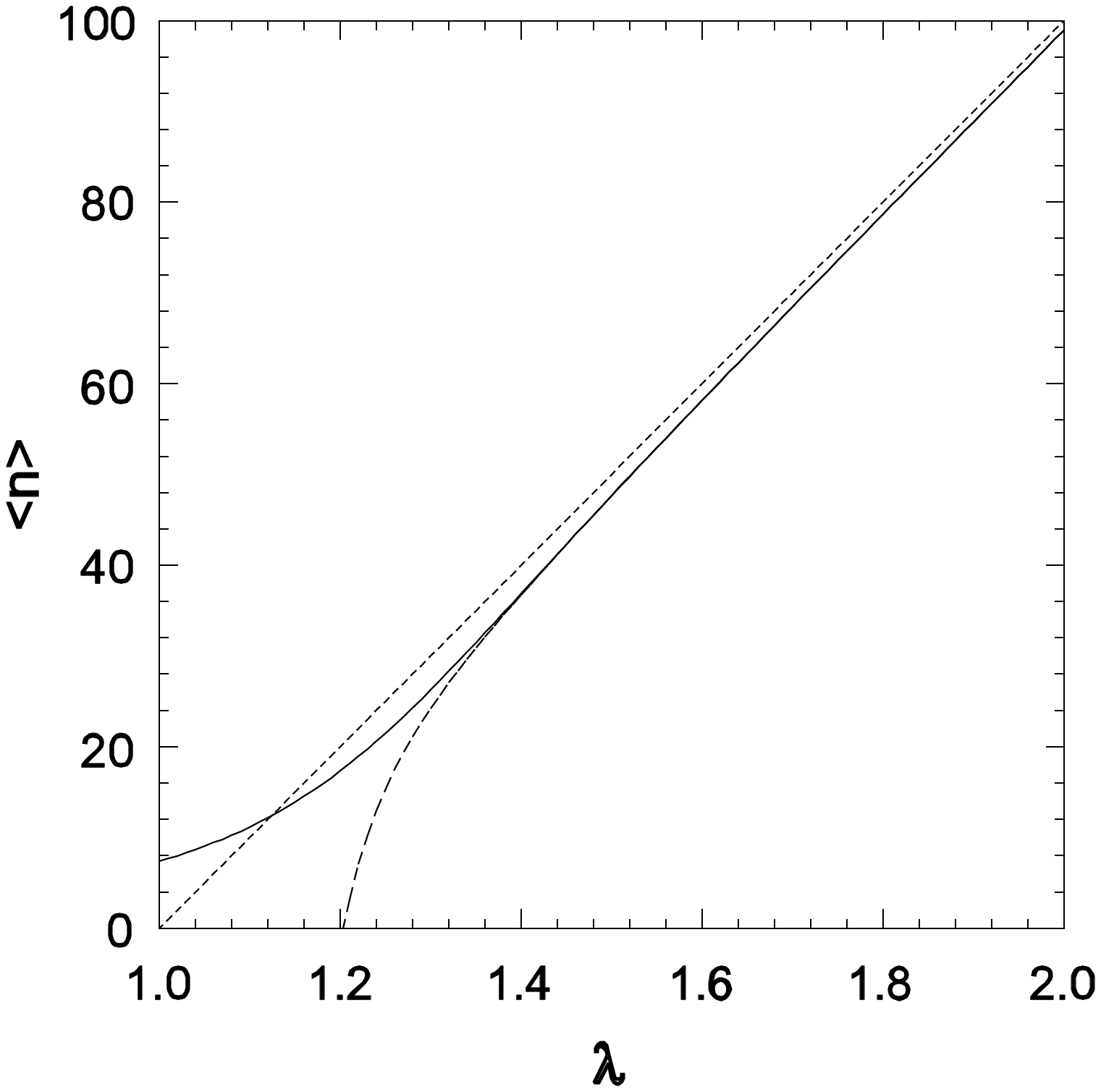}}
\label{c1f4}
\end{figure}
\begin{center}
{\sf Fig. 4. Stationary mean population size versus birth rate $\lambda$ in the
MVP with $\nu=0.01$.  Solid line: exact QS value; dotted line: mean-field prediction
$\overline{n}$; dashed line: series to order ${\cal O}(\epsilon^5)$.}
\end{center}

\begin{figure}[h]
\epsfysize=6cm
\epsfxsize=6cm
\centerline{
\epsfbox{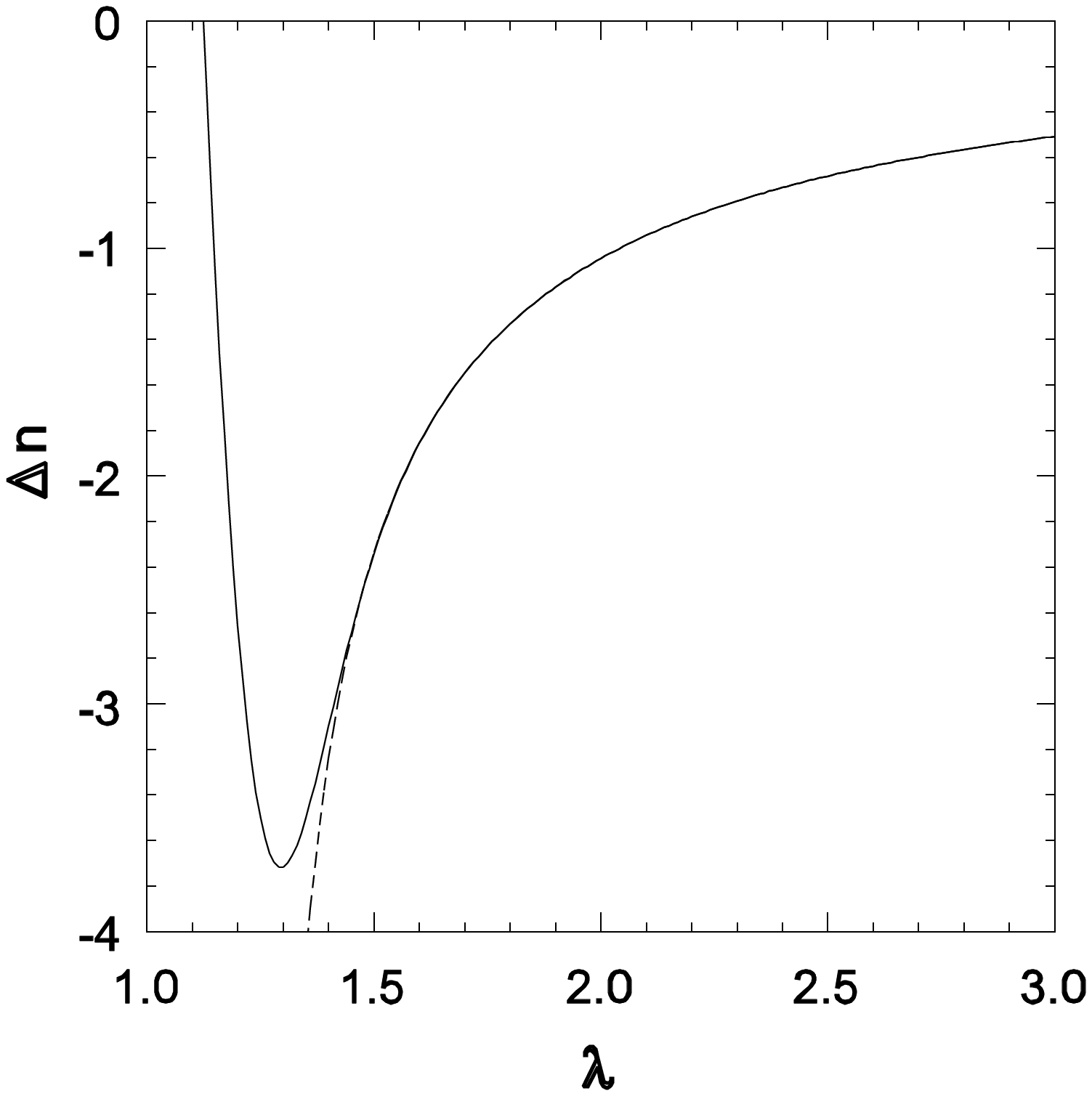}}
\label{c1f5}
\end{figure}
\begin{center}
{\sf Fig. 5. Difference $\Delta n$ between the stationary mean population
and the mean-field value in the MVP with $\nu=0.01$.
Solid line: exact QS value; dashed line: fifth-order series.}
\end{center}

\newpage
\begin{figure}[h]
\epsfysize=6cm
\epsfxsize=6cm
\centerline{
\epsfbox{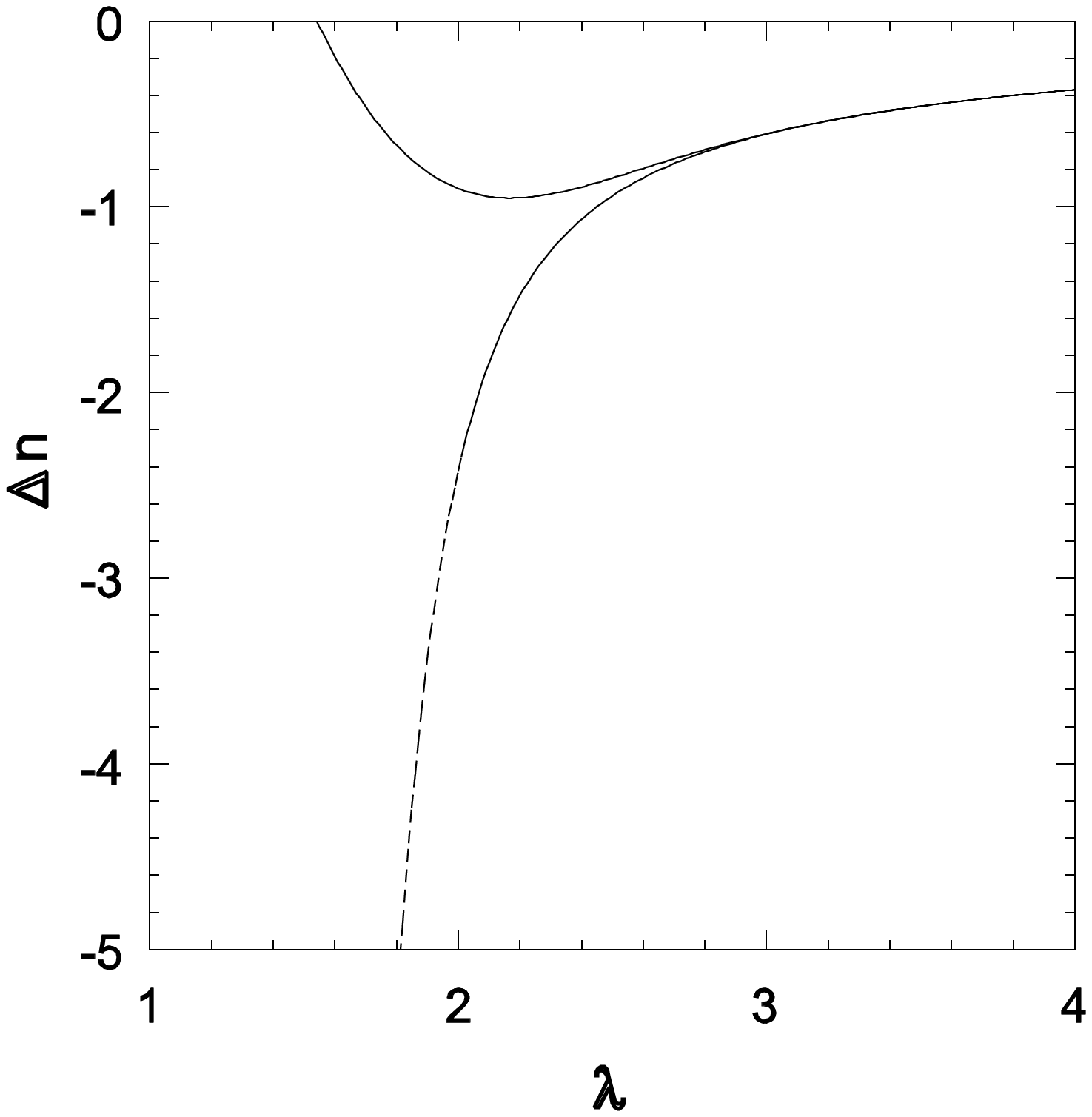}}
\label{c1f6}
\end{figure}
\begin{center}
{\sf Fig. 6. Same as Fig. 5 but for $\nu=0.1$.}
\end{center}

In order to gauge the importance of successive terms,
we plot (Fig. 8) $\Delta n$ (for $\nu=0.01$) as predicted
by the series to order $\epsilon^m$, for $m = 1$,..., 5.  For
$\lambda \geq 1.5$ very good agreement with the QS mean population
is obtained using the result to ${\cal O}(\epsilon^3)$.  For this
value of $\lambda$ the absolute difference between the three- and
five-term series is about 0.04 The relative difference is about one part
in a thousand; the difference rapidly decreases for larger
$\lambda$.

\begin{figure}[h]
\epsfysize=6cm
\epsfxsize=6cm
\centerline{
\epsfbox{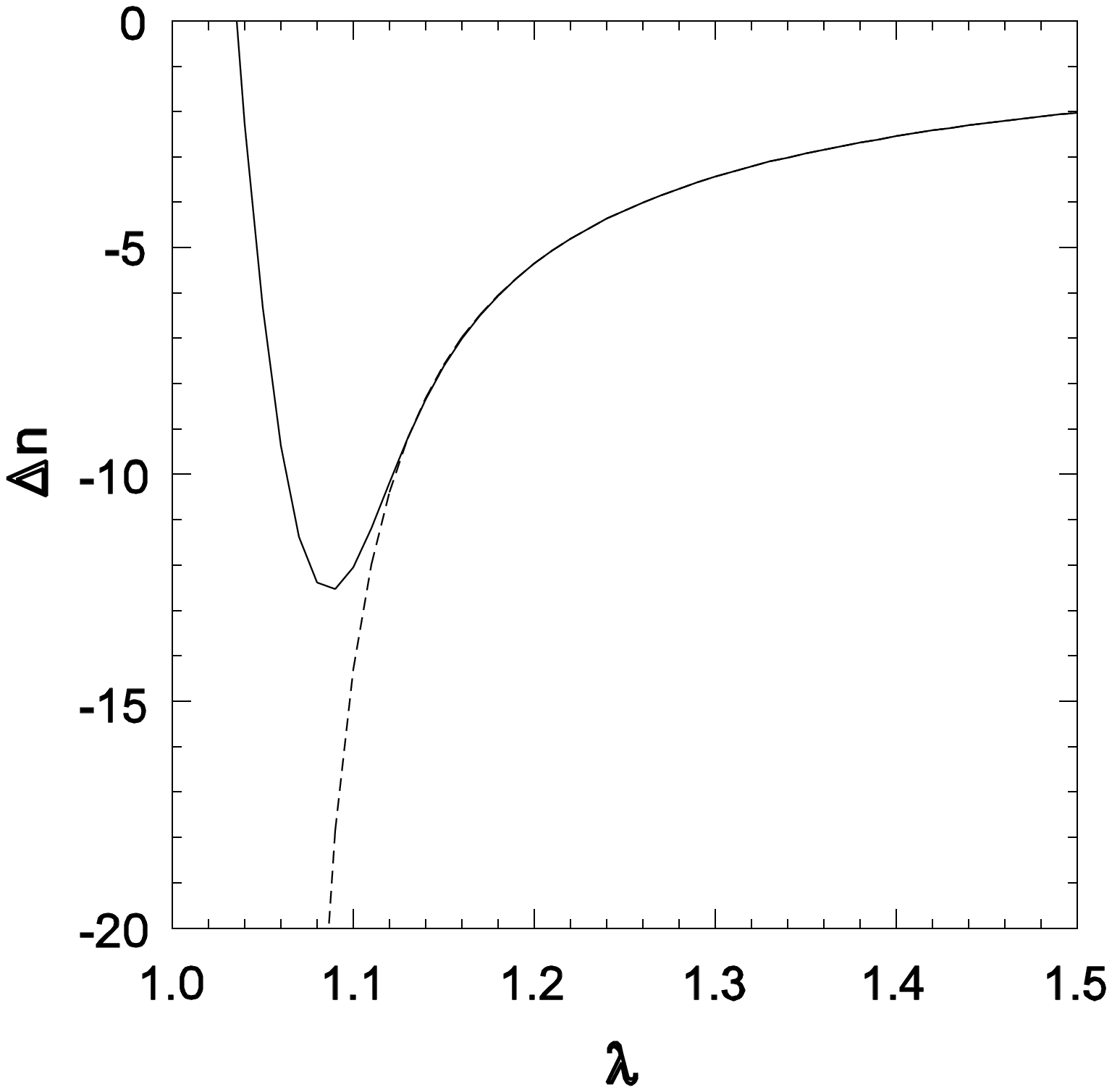}}
\label{c1f7}
\end{figure}
\begin{center}
{\sf Fig. 7. Same as Fig. 5 but for $\nu = 0.001$.}
\end{center}

\newpage

\begin{figure}[h]
\epsfysize=6cm
\epsfxsize=6cm
\centerline{
\epsfbox{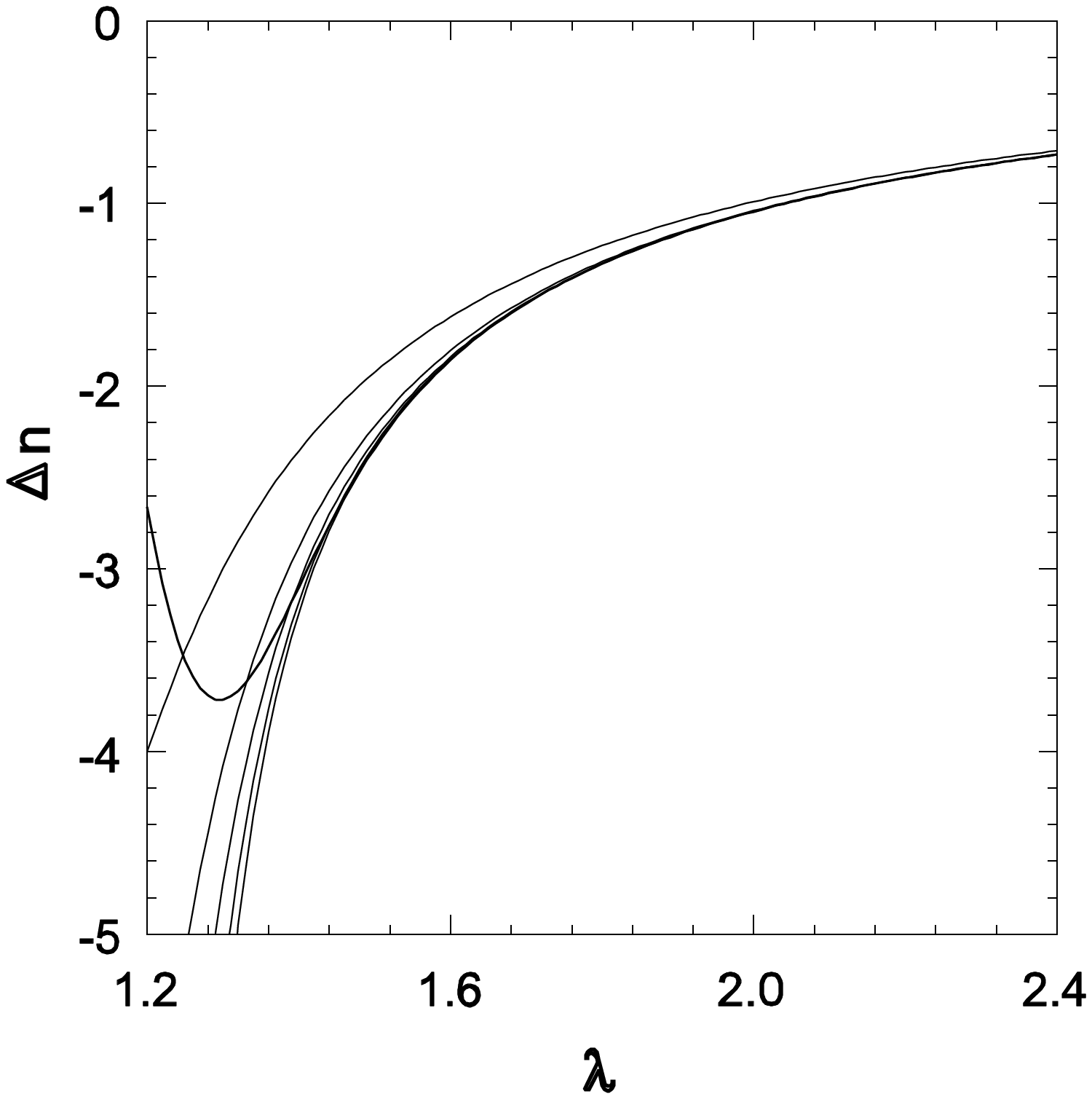}}
\label{c1f8}
\end{figure}
\begin{center}
{\sf Fig. 8. Stationary mean population size versus birth rate $\lambda$ in the
MVP with $\nu=0.01$.  The curve exhibiting a minimum near $\lambda = 1.3$
represents the exact QS value.  The other curves represent (in decreasing
magnitude) the series truncated at first, second,..., fifth order.}
\end{center}

\begin{figure}[h]
\epsfysize=6cm
\epsfxsize=6cm
\centerline{
\epsfbox{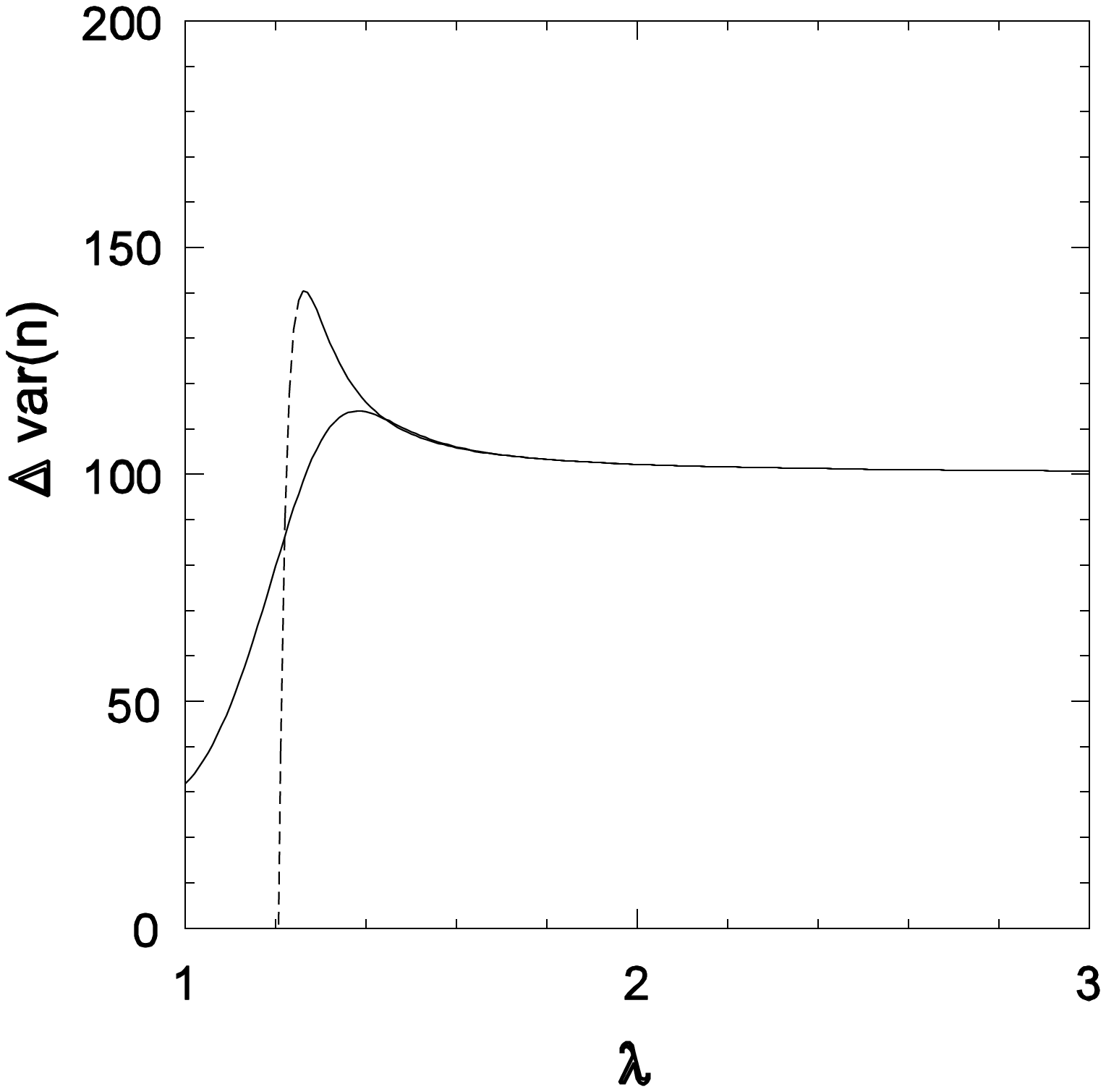}}
\label{c1f9}
\end{figure}
\begin{center}
{\sf Fig. 9. Difference $\Delta \mbox{var} (n)$ between the stationary variance
and its mean-field value in the MVP with $\nu=0.01$.
Solid line: exact QS value; dashed line: fifth-order series.}
\end{center}

\newpage
In light of the poor performance of the $\epsilon$ series in the
vicinity of $\lambda = 1$, it is natural to apply a resummation
technique such as Pad\'e approximants.  We therefore constructed the
[2,3], [3,2] and [4,1] approximants to the series $1 + g_1 \epsilon
+ \cdots + g_5 \epsilon^5$ and to the series for the logarithm of
this expression.  None of the Pad\'e approximants yielded any
improvement over the original series; the approximants are
ill-behaved (large and negative) near $\lambda = 1$ (they typically
exhibit a pole in this region) and reproduce the excellent agreement
with the QS results whenever the original series does.

\begin{figure}[h]
\epsfysize=6cm
\epsfxsize=6cm
\centerline{
\epsfbox{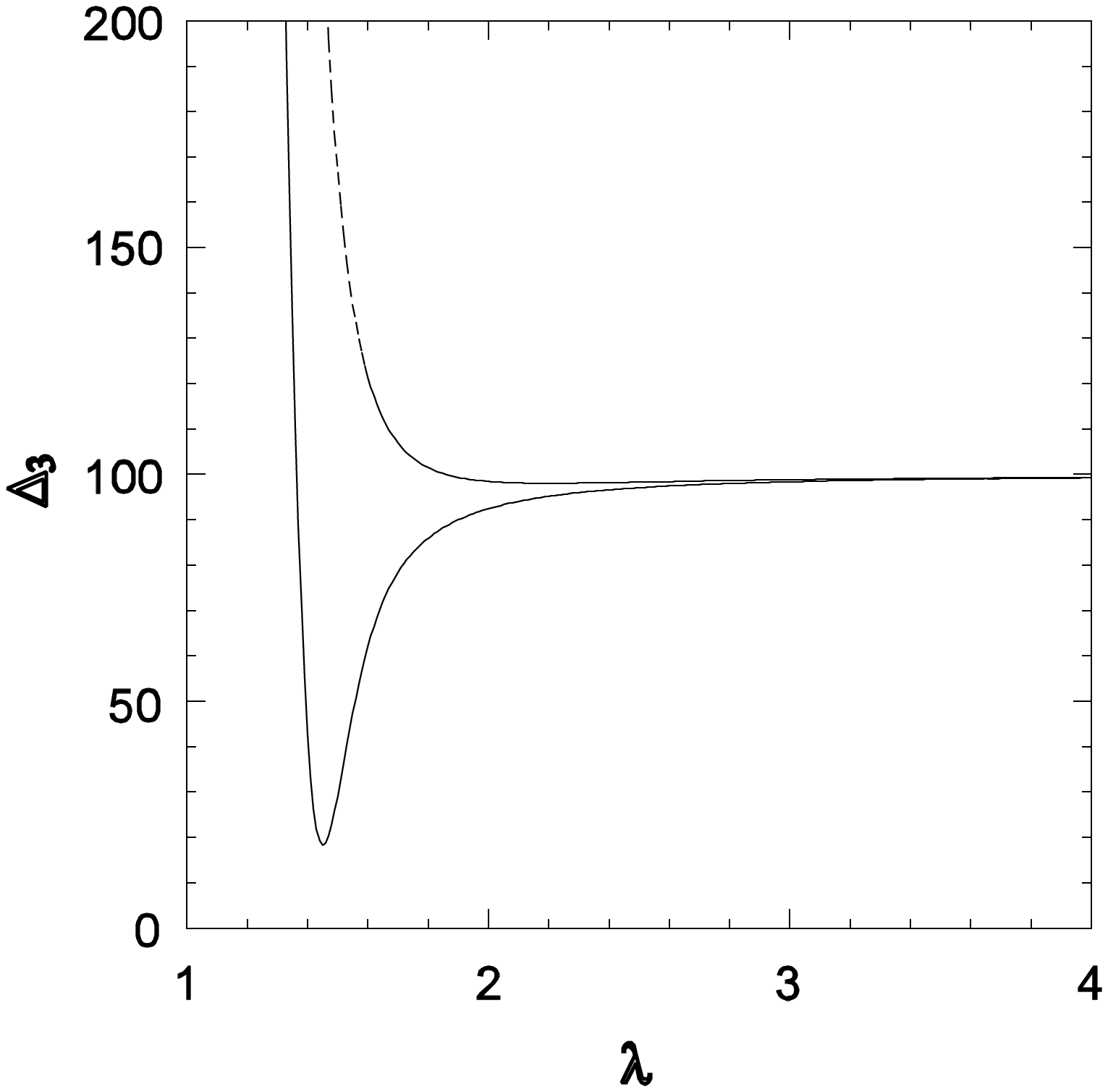}}
\label{c1f10}
\end{figure}
\begin{center}
{\sf Fig. 10. As in Fig. 9, but for the third central moment.}
\end{center}

The trends observed for the mean population continue when we compare
series and QS predictions for higher moments. In Fig. 9 we plot the
difference, $\Delta \mbox{var}(n)$, between the variance furnished
by these predictions and that given by mean-field theory. (Since the
latter yields a Poisson distribution, the variance is simply equal
to $\overline{n}$ in mean-field approximation.) For $\lambda \simeq
1$ the series prediction is useless; close agreement with the QS
result sets in (for $\nu = 0.01$), around $\lambda = 1.5$.  Note
that $\Delta \mbox{var}(n)$ approaches $1/\nu$ as $\lambda \to
\infty$, as predicted by Eq. (\ref{poidev}); the absolute value of the difference does not approach zero
for large $\lambda$, as it does for the mean. (The {\it relative}
difference $\Delta \mbox{var}(n)/\mbox{var}(n)$ does of course
approach zero in the limit $\lambda \to \infty$.)  In Fig. 10 we
compare the difference from mean-field theory for the third central
moment $\langle ( n - \langle n \rangle )^3 \rangle$ (mean-field
theory yields $\overline{n}$ for this quantity); Fig. 11 is a
similar comparison for the fourth central moment [in this case the
mean-field prediction is $(3\overline{n} +1)  \overline{n}$].  The
series agrees well with the QS result (again, for the case $\nu =
0.01$), for $\lambda \geq 2.5$.

\begin{figure}[h]
\epsfysize=6cm
\epsfxsize=6cm
\centerline{
\epsfbox{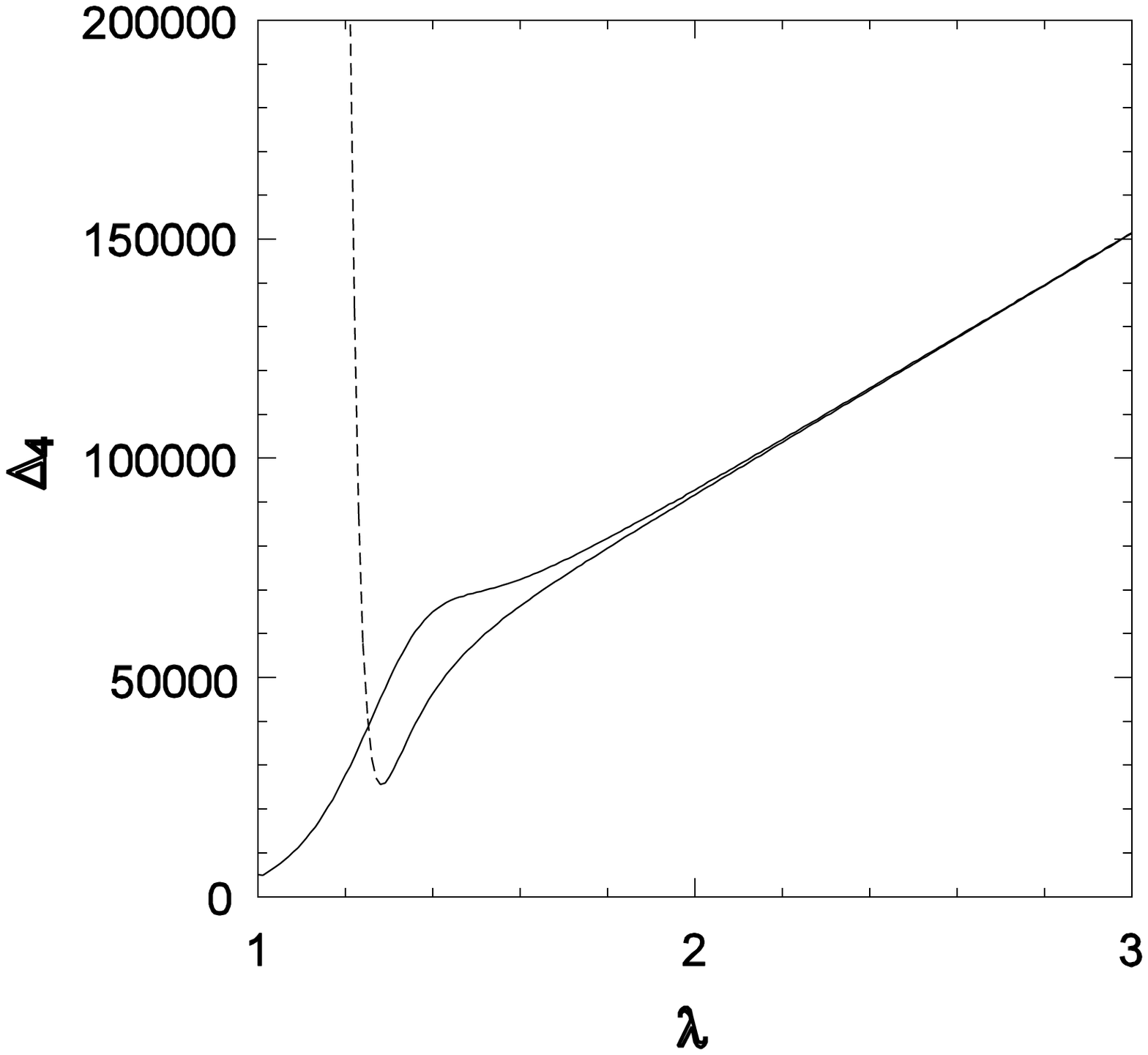}}
\label{c1f11}
\end{figure}
\begin{center}
{\sf Fig. 11. As in Fig. 9, but for the fourth central moment.}
\end{center}

On the basis of these comparisons we conclude that the perturbation
theory developed in the previous section is incapable of describing
the vicinity of the transition ($\lambda \approx 1$), but agrees
extremely well with quasi-stationary properties (including higher
moments) above a certain, not very large value of $\lambda$. In this
regime, the perturbation approach yields accurate analytic
expressions for QS properties.

\begin{figure}[h]
\epsfysize=6cm \epsfxsize=6cm \centerline{ \epsfbox{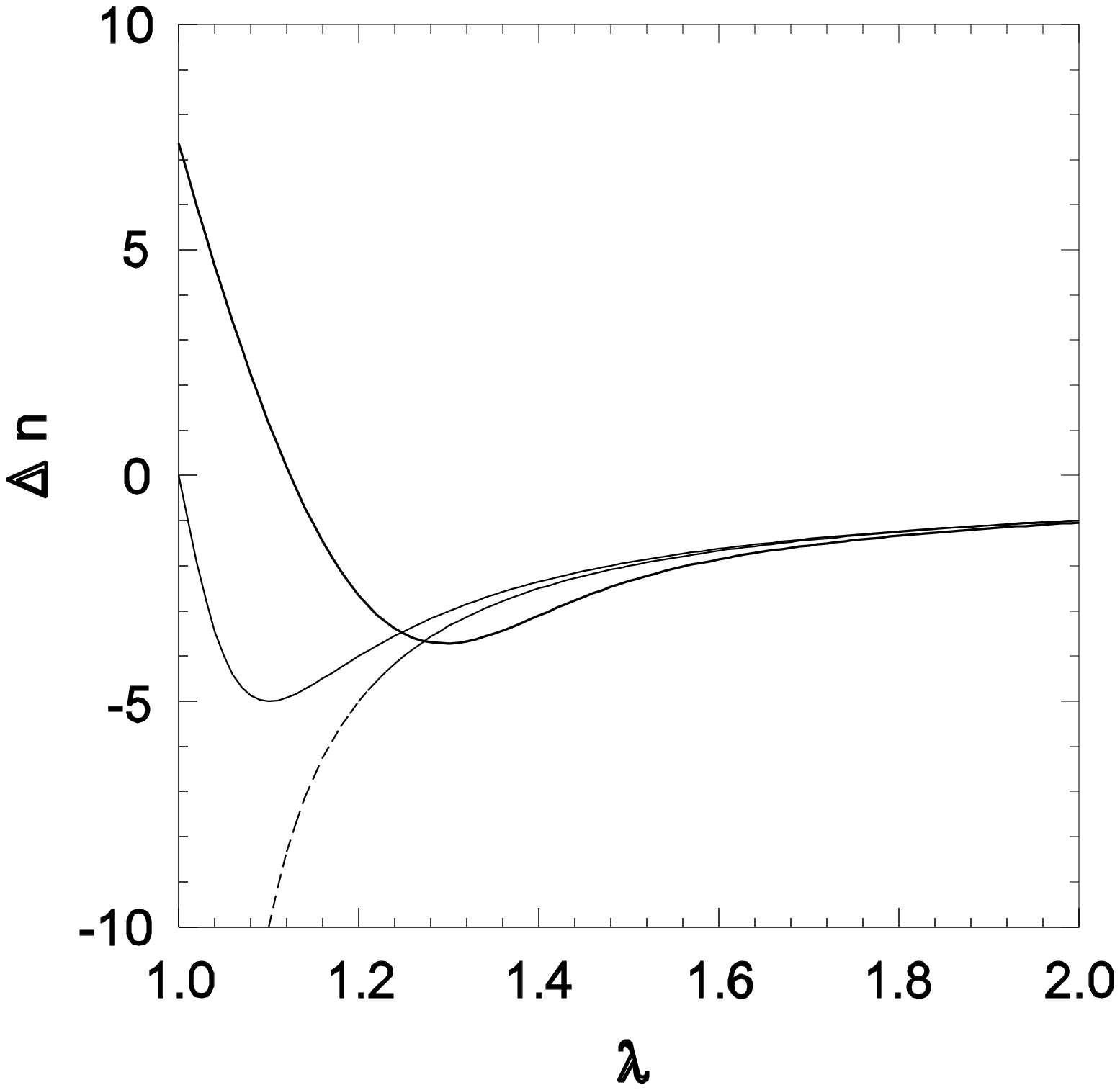}}
\label{c1f12}
\end{figure}
{\sf Fig. 12. Correction $\Delta n$ to the mean population size
versus birth rate $\lambda$ in the MVP with $\nu=0.01$. Bold curve:
QS distribution; dotted line: lowest order correction in
$\Omega$-expansion; solid line: $\epsilon$ series truncated at first
order.}

In Fig. 12 we compare the lowest order $\Omega$-expansion prediction
for $\Delta n$ and the corresponding result of the $\epsilon$ series
(truncated at order $\epsilon$) with the QS result, for $\nu=0.01$.
The $\Omega$-expansion result is seen to be slightly better,
although it diverges as $\lambda \to 1$.  Fig. 13 is a similar
comparison for $\Delta$var$(n)$.  The $\Omega$-expansion is again
slightly superior. Finally, Fig. 14 shows that the fifth-order
$\epsilon$ series is superior to the second order $\Omega$-expansion
result, as is to be expected.

\begin{figure}[h]
\epsfysize=6cm \epsfxsize=6cm \centerline{ \epsfbox{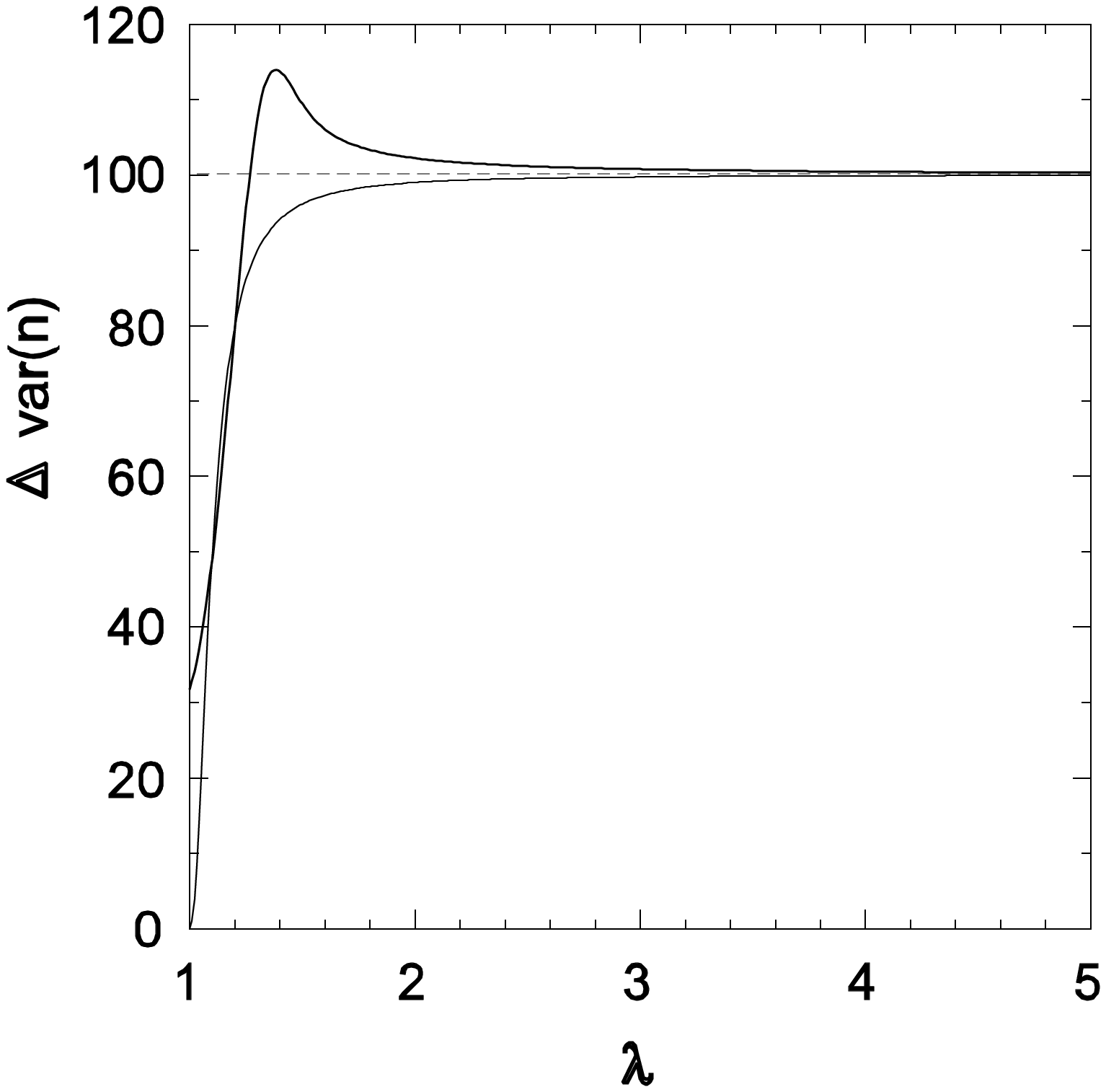}}
\label{c1f13}
\end{figure}
{\sf Fig. 13. Correction $\Delta$var$(n)$ versus birth rate
$\lambda$ in the MVP with $\nu=0.01$. Symbols as in Fig. 12.}

\begin{figure}[h]
\epsfysize=6cm \epsfxsize=6cm \centerline{ \epsfbox{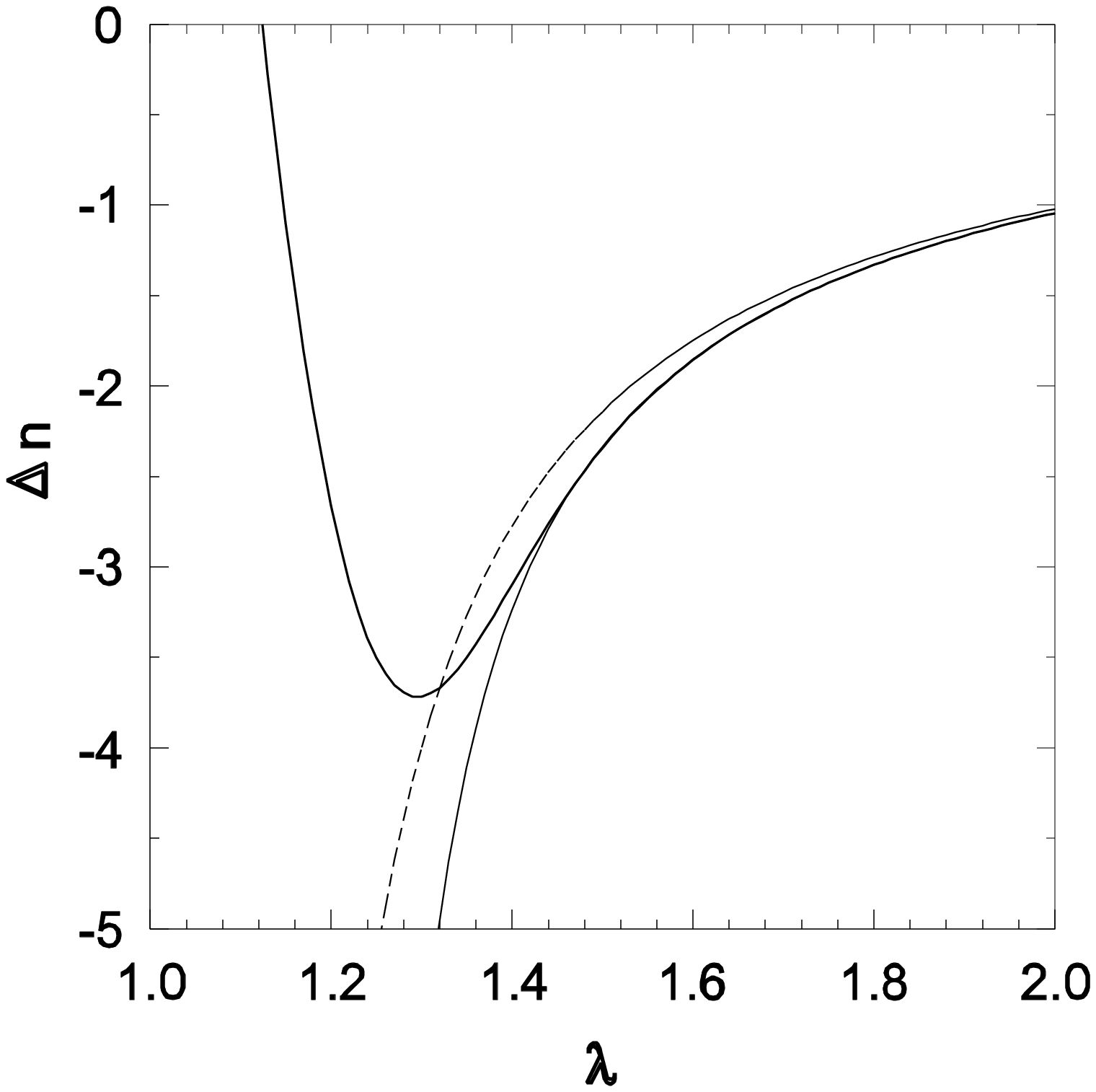}}
\label{c1f14}
\end{figure}
{\sf Fig. 14. Correction $\Delta n$ to the mean population size
versus birth rate $\lambda$ in the MVP with $\nu=0.01$. Symbols as
in Fig. 12.}

\section{MVP with input}

In this section we examine the effect of a steady input of
individuals at rate $\gamma$.  For $\gamma > 0$ the $n=0$ state is
no longer absorbing, and the system approaches a true stationary
state as $t \to \infty$.  The forward transition rate is now
$w_{n,n+1} = \gamma + \lambda n$; the reverse rate $w_{n-1,n} = n +
\nu n(n-1) $ as before.  The evolution operator is that of the MV
process with the addition of a term $\gamma (\pi -1)$, in the
notation of \cite{pert}.  This corresponds to a new term in the
action (i.e., the argument of the exponential in Eq. (\ref{F})),
\begin{equation}
\gamma \int_0^t (i \psi' -1) dt'= \gamma \int_0^t
\hat{\psi} dt'
\label{newterm}
\end{equation}

We now proceed as in Sec. II, introducing the shift of variable
$\psi = \overline{n} + \phi(t)$.  Equating the coefficient of
$\hat{\psi}$ in the action to zero yields the stationary solution
of the macroscopic equation, that is,
\begin{equation}
\overline{n} = \frac{1}{2 \nu} (\lambda - 1 + w)
\label{newnbar}
\end{equation}
where $w = \sqrt{(\lambda-1)^2 + 4 \gamma \nu}$.
After the shift, the action has the same form as for the MV process
without a source, but the factors associated with the source
and the bifurcation are changed.  The factor for the source is now:
\begin{equation}
 \frac{1}{2} \overline{n} (\lambda + 1 - w) \equiv \omega \overline{n}
\label{newsource}
\end{equation}
while that for the bifurcation is $1-w$.  The factors associated with
the conjunction and the 4-vertex are $-\nu$, as before.

Consider the contribution to $\langle n_\infty \rangle/\overline{n}$
due to the lowest order diagram (i.e., the
second diagram of Fig. 2);
this is readily seen to be  $- \omega \nu/w^2$.  As before, we may dress the
diagram (see Fig. 3), which again has the effect of multiplying the above
result by the factor $\kappa = [1+ \nu/w^2]^{-1}$.  Letting
$\epsilon = 1-\kappa$, the lowest order contribution, when dressed,
is $-\omega \epsilon$.  For diagrams with $n \geq 4$ both rightmost
source and the leftmost conjunction can be dressed, leading to
the algebraic factor
\begin{equation}
F_{alg} = (-1)^{c+f} u^b \frac{\omega^s}{w^n}
 \left(\frac{\lambda - 1+w}{2}\right)^{s-1}
 \frac{\nu^m}{(1+\nu/w^2)^2}.
\label{falgs}
\end{equation}
where $m=n-c$ and $u=1-w$ as before.  In terms of the parameter $\epsilon$
we have
\begin{equation}
F_{alg} = (-1)^{c+f} h u^b v^s  w^{2m-n}
 \frac{\epsilon^m}{(1-\epsilon)^{m-2}}
\label{falgse}
\end{equation}
where
\begin{equation}
h = \frac{2}{\lambda + w - 1}
\end{equation}
and
\begin{equation}
v =  \frac{\lambda^2 - u^2}{4} .
\end{equation}
With these results the enumerations derived for the original
process may be used to develop an $\epsilon$ series for the
MVP with input.  The input rate $\gamma$ enters via the expression
for $w$.  Up to third order we find,
\begin{eqnarray}
\nonumber
\langle n_\infty \rangle &=& \overline{n} \left[1 - \omega \epsilon
+ vh \left(2uw - 5\frac{v}{w} \right) \epsilon^2 \right.
\\
&+& \left. vh \left( -4uw - 10 u^2 + 19 v + 55u \frac{v}{w}
-60 \left(\frac{v}{w} \right)^2  \right) \epsilon^3 \right]
\label{expsi}
\end{eqnarray}

Series is compared with the exact value of $\langle n_\infty
\rangle$ (from the stationary solution of the master equation,
obtained numerically), and with the mean-field prediction
$\overline{n}$ in Fig. 15, for parameters $\gamma = \mu = 0.01$.
Excellent agreement between series and the exact result is found for
$\lambda \geq 1.55$ and again (see inset) for $\lambda \leq 0.75$.
In the regions of agreement $\epsilon$ is small ($\epsilon \leq
0.14$ for $\lambda \leq 0.75$; $\epsilon \leq 0.032$ for $\lambda
\geq 1.55$), but it becomes large in the intermediate region,
attaining a maximum of about 0.96 for $\lambda = 1$. In this region
the series prediction deviates strongly from the true value, and can
become negative. Similar results are found for other values of
$\gamma$ and $\nu$.

\begin{figure}[h]
\epsfysize=6cm \epsfxsize=6cm \centerline{ \epsfbox{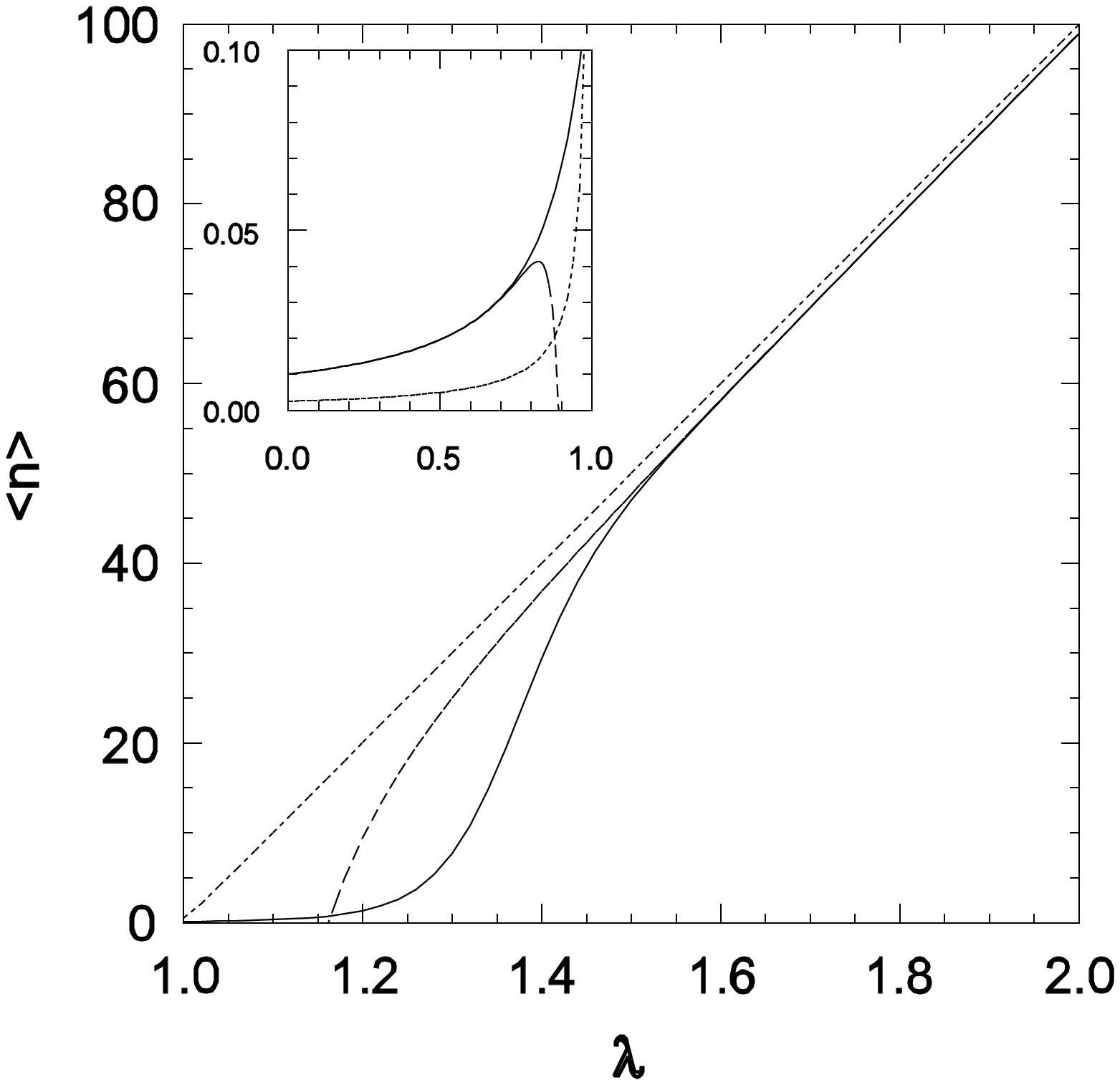}}
\label{c1f15}
\end{figure}
{\sf Fig. 15. Stationary mean population size versus birth rate
$\lambda$ in the MVP with $\nu=0.01$ and input at rate
$\gamma=0.01$.  Symbols as in Fig. 4.}

\section{Conclusions}

We apply the path-integral based perturbation theory for Markovain
birth-and-death processes \cite{peliti85,pert} to the
Malthus-Verhulst process and a variant of this process that includes
particle input.  At zeroth order the formalism yields a Poisson
distribution whose mean is given by the mean-field or macroscopic
equation. Computational enumeration of diagrams allows us to derive
the series coefficients for the first four stationary moments of the
process. The expansion parameter, $\epsilon$, is small when the mean
population size $\langle n \rangle$ is large, but is of order unity
when $\langle n \rangle$ is small.  In the latter regime the
expansion fails, but in the former its predictions are in near-perfect
agreement with numerical results, despite evidence that the series
is divergent.  Truncating the series at third order, we find excellent
agreement with numerical results when the mean population
$\langle n \rangle \geq 35$.  Our analysis yields asymptotic expressions
(valid for large population size) for the moments of MVP.

Our study of the MVP with input shows that the poor behavior of the
series, when $\epsilon$ is not small, is not due
the presence of an absorbing state, since input eliminates such a
state from the process.  Since the present approach is based on
systematic approximations to mean-field theory, one should not
expect it to yield useful results where the latter is seriously in
error, as is the case for $\lambda \leq 1$ in the MVP with or
without input.  For parameter values such that the mean-field solution is
reasonable, the series provides useful corrections to it.  Development
of a globally accurate perturbation method remains as an open challenge,
one we hope to explore in future work.

Comparison with van Kampen's $\Omega$-expansion shows that (for the
problems considered here), the latter method, to lowest order,
yields results that are very similar, though slightly superior to our
first-order expansion.  Including higher order terms, the quality of
the $\epsilon$ series improves.  In the present case (and in other
problems likely to arise in stochastic modeling), deriving
higher-order corrections is simpler using the diagrammatic approach,
making our method an attractive alternative to the
$\Omega$-expansion.

\noindent{\bf Acknowledgments}

NRM is grateful to FAPEMIG for financial support, and to the
hospitality of the Universidade Federal de Minas Gerais. This work
was supported by FAPEMIG and CNPq, Brazil.




\end{document}